\documentclass[iop]{emulateapj-rtx4}
\usepackage{apjfonts}

\input psfig.tex

\newcommand{\kms}{km\,s$^{-1}$} 
\newcommand{\Msun}{M$_{\odot}$}

\newcommand{\picplace}[1]{\vbox{\hrule\@height 0.4pt\@width\hsize
\hbox to\hsize{\vrule\@width 0.4pt\@height#1\hfil
\vrule\@width 0.4pt\@height#1}\hrule\@height 0.4pt\@width\hsize}}

\slugcomment{Accepted for Publication in ApJ}

\shorttitle{The Nature of Extended MSTOs in LMC Star Clusters}
\shortauthors{Goudfrooij et al.}

\begin{document}


\title{Population Parameters of Intermediate-Age Star Clusters in the
  Large Magellanic Cloud. \\ III. Dynamical Evidence for a Range of Ages Being
  Responsible for Extended Main Sequence Turnoffs\altaffilmark{1}}  


\author{Paul Goudfrooij\altaffilmark{2}, Thomas H. Puzia\altaffilmark{3},
  Rupali Chandar\altaffilmark{4}, and Vera Kozhurina-Platais\altaffilmark{2}}

\altaffiltext{1}{Based on observations with the NASA/ESA {\it Hubble
    Space Telescope}, obtained at the Space Telescope Science
  Institute, which is operated by the Association of Universities for
  Research in Astronomy, Inc., under NASA contract NAS5-26555} 
\altaffiltext{2}{Space Telescope Science Institute, 3700 San Martin
  Drive, Baltimore, MD 21218; goudfroo@stsci.edu, verap@stsci.edu} 
\altaffiltext{3}{Department of Astronomy and Astrophysics, Pontificia
  Universidad Cat\'olica de Chile, Av. Vicu\~{n}a Mackenna 4860, Macul
  7820436, Santiago, Chile; tpuzia@gmail.com} 
\altaffiltext{4}{Department of Physics and Astronomy, The University of Toledo,
  2801 West Bancroft Street, Toledo, OH 43606; rupali.chandar@utoledo.edu} 











\begin{abstract}
We present new analysis of 11 intermediate-age (1-2~Gyr) star clusters
in the Large Magellanic Cloud based on Hubble Space Telescope imaging data.
Seven of the clusters feature main sequence turnoff (MSTO)
regions that are wider than can be accounted 
for by a simple stellar population, whereas their red giant
branches indicate a single value of [Fe/H]. The star clusters cover a
range in present-day mass from about $1\times 10^4$ \Msun\ to $2
\times 10^5$ \Msun.   
We compare radial distributions of stars in the upper and lower parts
of the MSTO region, and calculate cluster masses and escape velocities
from the present time back to a cluster age of 10 Myr. 
Our main result is that for all clusters in our sample with estimated
escape velocities $v_{\rm esc} \ga 15$ \kms\ at an age of 10 Myr,
the stars in the brightest half of the MSTO region are significantly
more centrally concentrated than the stars in the faintest half {\it
  and\/} more massive red giant branch and asymptotic giant branch
stars. This is not the case for clusters with $v_{\rm esc} \la 10$
\kms\ at an age of 10 Myr. We argue that the wide MSTO region of such
clusters is mainly caused by to a $\sim200 - 500$~Myr range in the ages 
of cluster stars due to extended star formation within the cluster
from material shed by first-generation stars featuring slow stellar 
winds. Dilution of this enriched material by accretion of ambient
interstellar matter is deemed plausible if the spread of
[Fe/H] in this ambient gas was very small when the second-generation
stars were formed in the cluster.    
\end{abstract}




\section{Introduction}              \label{s:intro}

For several decades, the standard paradigm for globular clusters (GCs) was
that they consist of stars born at the same time out of the same
material. This scenario has faced serious challenges over the last
decade. It is now known that the most massive GCs in our Galaxy such as
$\omega$\,Cen and M\,54 host multiple red giant branches (RGBs) due to
populations with different [Fe/H]
\citep[e.g.,][]{sarlay95,lee+99,hilric00,vill+07,carr+10}. Somewhat less 
massive Galactic GCs such as NGC~2808, NGC~1851 and 47~Tuc show multiple
sub-giant branches (SGBs) and/or multiple main sequences, which are typically
interpreted as populations with different Helium abundances  
\citep[e.g.,][]{piot+07,milo+08,ande+09}. While lower-mass Galactic GCs
typically do not show clear evidence for multiple populations from optical
broad-band photometry, spectroscopic surveys do show that {\it light
elements\/} such as C, N, O, F, and Na show significant star-to-star abundance
variations (often dubbed ``Na-O anticorrelations'') within all Galactic GCs
studied to date in sufficient detail (\citealt{carr+09}, and references
therein). Since these abundance variations have been found among 
RGB stars as well as main sequence (MS) stars within a given GC \citep{grat+04}, the
suggested cause of the variations is that secondary generation(s) of stars
formed out of material shed by an older, evolved population within the cluster.
While the chemical processes responsible for causing the light element abundance
variations  have largely been identified as proton-capture reactions in
Hydrogen burning at high temperature \citep[$\ga 40 \times 10^6$ K, see
e.g.][]{grat+04}, the old age of Galactic GCs has precluded a clear picture of
the time scales and hence the types of stars involved in the chemical
enrichment of the second-generation stars. Currently, the most popular
candidates are {\it (i)\/} intermediate-mass AGB stars ($4 \la
{\cal{M}}/M_{\odot} \la 8$, hereafter IM-AGB; e.g., \citealt{danven07} and
references therein), {\it (ii)\/} rapidly rotating massive stars (often
referred to as ``FRMS''; e.g., \citealt{decr+07}) and {\it (iii)\/} massive
binary stars \citep{demink+09}.  

Recently, deep CMDs from images taken with the Advanced Camera for Surveys
(ACS) aboard the Hubble Space Telescope (HST) provided conclusive evidence
that several massive intermediate-age star clusters in the Magellanic Clouds
host extended and/or multiple main sequence turn-off (MSTO) regions  
(\citealt{mack+08,glat+08,milo+09,goud+09}, hereafter Paper I; 
\citealt{goud+11a}, hereafter Paper II), in some cases accompanied by
composite red clumps \citep{gira+09,rube+11}. To date, these observed properties have
been interpreted in three main ways: {\it (i)\/} Bimodal age distributions
\citep{mack+08,milo+09}, {\it (ii)\/} age spreads of 200\,--\,500 Myr
(Paper II; \citealt{gira+09,rube+10,rube+11}), and {\it (iii)\/} spreads in
rotation velocity among turn-off stars (\citealt{basdem09}). 

In this third paper of this series we study the dynamical properties of 11
intermediate-age star clusters in the LMC: seven clusters that exhibit
extended MSTO regions (hereafter eMSTOs) and four that do not. 
We determine and compare radial distributions of cluster stars at
different evolutionary phases and evaluate the evolution of the
clusters' masses and escape velocities from an age of 10 Myr to their
current age. This analysis reveals new findings relevant to the
assembly of these intermediate-age star clusters and their
evolutionary association with multiple stellar populations in ancient
Galactic globular clusters.    

The remainder of this paper is organized as follows. \S\ 
\ref{s:sample} presents the cluster sample. In \S\ \ref{s:rad_dist} we
determine the radial distributions of stars in various different
evolutionary phases and study the dependencies of these distributions with
cluster escape velocities as a function of time. \S\ \ref{s:disc}
uses our new results to constrain the origin of 
multiple populations in these clusters, 
and \S\ \ref{s:conc} presents our main conclusions. 

\section{Target Clusters} \label{s:sample}

Our main sample of intermediate-age clusters is that presented in Paper
II. All these clusters were observed within one HST/ACS program using the same
observational setup, using both short and long exposures to yield high-quality 
photometry throughout the CMDs. The sample consists of star clusters in the LMC
with integrated {\it UBV\/} colors consistent with SWB \citep{swb80} parameter
values in the  range IV{\sc b}~$-$~VI, which translates to estimated ages
between roughly 1.0 and 2.5 Gyr. This turned out to be fully consistent with
the ages actually found from isochrone fitting. The main properties of these star
clusters, all of which were found to host eMSTO regions, are listed in
Table~\ref{t:sample}.  For comparison with clusters in the same
age range that do {\it not\/} contain eMSTO regions, we also use the four LMC
clusters NGC~1644, NGC~1652, NGC~1795, and IC~2146 that were studied by
\citet{milo+09} using HST/ACS photometry. To our knowledge, these are the only
four clusters in the LMC in the age range 1\,--\,2 Gyr that are known to date
{\it not\/} to harbor eMSTO regions \citep{milo+09} from high-quality
photometry at HST resolution.  

\begin{table*}[tbh]
\begin{center}
\footnotesize
\caption{Main properties of the eMSTO star clusters studied in this paper.}
 \label{t:sample}
\begin{tabular}{@{}lcccccc@{}}
\multicolumn{3}{c}{~} \\ [-2.5ex]   
 \tableline \tableline
\multicolumn{3}{c}{~} \\ [-1.8ex] 
\multicolumn{1}{c}{Cluster} & $V$ & Ref. & SWB & Age & [Z/H] & $A_V$ \\
\multicolumn{1}{c}{(1)}     & (2) & (3)  & (4) & (5) & (6)   & (7)  
 \\ [0.5ex] \tableline  
\multicolumn{3}{c}{~} \\ [-2.ex]              
NGC 1751 & 11.67 $\pm$ 0.13 & 1 & VI & $1.40 \pm 0.10$ & $-0.3 \pm 0.1$ & 
 $0.40 \pm 0.01$ \\
NGC 1783 & 10.39 $\pm$ 0.03 & 1 & V  & $1.70 \pm 0.10$ & $-0.3 \pm 0.1$ & 
 $0.02 \pm 0.02$ \\
NGC 1806 & 11.00 $\pm$ 0.05 & 1 & V  & $1.67 \pm 0.10$ & $-0.3 \pm 0.1$ & 
 $0.05 \pm 0.01$ \\
NGC 1846 & 10.68 $\pm$ 0.20 & 1 & VI & $1.75 \pm 0.10$ & $-0.3 \pm 0.1$ & 
 $0.08 \pm 0.01$ \\
NGC 1987 & 11.74 $\pm$ 0.09 & 1 & {\sc IVb} & $1.05 \pm 0.05$ & $-0.3\pm0.1$ & 
 $0.16 \pm 0.04$ \\
NGC 2108 & 12.32 $\pm$ 0.04 & 2\tablenotemark{a} & {\sc IVb} & $1.00\pm0.05$ & 
 $-0.3\pm0.1$ & $0.50 \pm 0.03$ \\
  LW 431 & 13.67 $\pm$ 0.04 & 2\tablenotemark{a} & VI & $1.75\pm0.10$ & 
 $-0.3\pm0.1$ & $0.15\pm0.01$ \\ [0.5ex] \tableline
\multicolumn{3}{c}{~} \\ [-2.5ex]              
\end{tabular}
\tablecomments{Column (1): Name of star cluster. (2): Integrated $V$
  magnitude. (3): Reference of $V$ magnitude. Ref. 1: \citet{goud+06}; Ref.\
  2: \citet{bica+96}. (4): SWB type from \citet{bica+96}. (5) Age in Gyr from
  Paper II. (6) Metallicity from paper II. (7) $A_V$ from Paper II.}  
\tablenotetext{a}{uncertainty only includes internal errors associated with
  measurements of cluster and one background aperture.}
\end{center}
\end{table*}

\section{Radial Distributions of Stars in Different Regions of the CMD} 
\label{s:rad_dist} 

\subsection{Motivation}

  Since the eMSTO regions in our target clusters may be due to the
  presence of more than one simple stellar population, it is important
  to find out whether different subregions of the eMSTO have
intrinsically different radial distributions, i.e., differences beyond those
that can be expected for simple (coeval) stellar populations due to dynamical
evolution. A well-known example of the latter is mass segregation due to
dynamical friction which slows down  stars on a time scale that is inversely
proportional to the mass of the star \citep[e.g.,][]{sasl85,spit87,mayheg97}
so that massive stars have a more centrally concentrated distribution over
time than less massive ones.  Another reason for studying the radial
distributions of stars in relevant evolutionary phases in the CMD
is that the masses and ages of these clusters are such
that the two-body relaxation times of stars within the clusters are comparable  
to the cluster ages: The mean two-body relaxation time $t_{\rm relax}$ of
stars within a cluster's half-mass radius is 
\begin{equation}
t_{\rm relax} \approx \frac{N}{8\,\ln N} \: t_{\rm cross} 
 \approx \frac{N}{8\,\ln N} \; r_{\rm h}^{1.5} \; (G\,{\cal{M}}_{\rm cl})^{-0.5} 
\end{equation}
\citep{bintre87} where $t_{\rm cross}$ is the crossing time at the half-mass
radius, $N$ is the number of stars in the cluster, $r_h$ is the half-mass
radius, $G$ is the gravitational constant, and ${\cal{M}}_{\rm cl}$ is the cluster
mass. $t_{\rm relax}$ values for the clusters in our sample are listed in
Table~\ref{t:dynamics}. 
For a typical cluster among the 50\% most massive clusters in our sample, $N
\simeq 1.5 \times 10^5$, ${\cal{M}}_{\rm cl} \simeq 10^5$\,\Msun, and $r_{\rm h} \simeq
8$ pc, resulting in $t_{\rm relax} \simeq 2$ Gyr.  Hence, dynamical imprints from the
formation epoch of the stars within these clusters may well be still
observable, in contrast to the situation in the vast majority of
Galactic globular clusters ($\omega$\,Cen is the exception, see e.g.\
\citealt{bell+09}). This renders these intermediate-age clusters excellent
probes of the nature of multiple stellar populations in star
clusters. 

\subsection{Measurements and Implications}

\begin{figure}[tb]
\centerline{
\psfig{figure=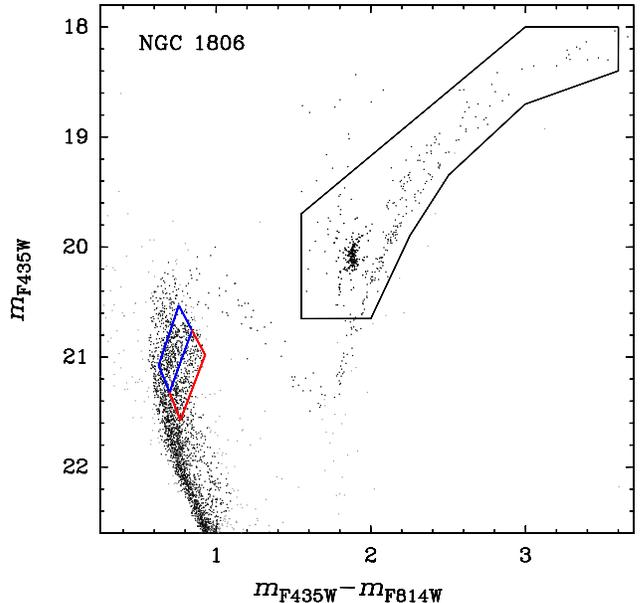,width=8.3cm}
}
\caption{Illustration of the three regions used to derive the radial surface number density
  distributions shown in Fig.\ \ref{f:rad_dists}, superposed onto the CMD of
  NGC 1806 as shown in the middle panel of Fig.\ 2 of Paper II. The
  ``upper MSTO region'' is outlined by blue solid lines, the ``lower MSTO
  region'' by red lines,  and the ``RC, RGB, and AGB'' by black lines. 
\label{f:regions}}
\end{figure}

\begin{figure*}[tbp]
\centerline{
\psfig{figure=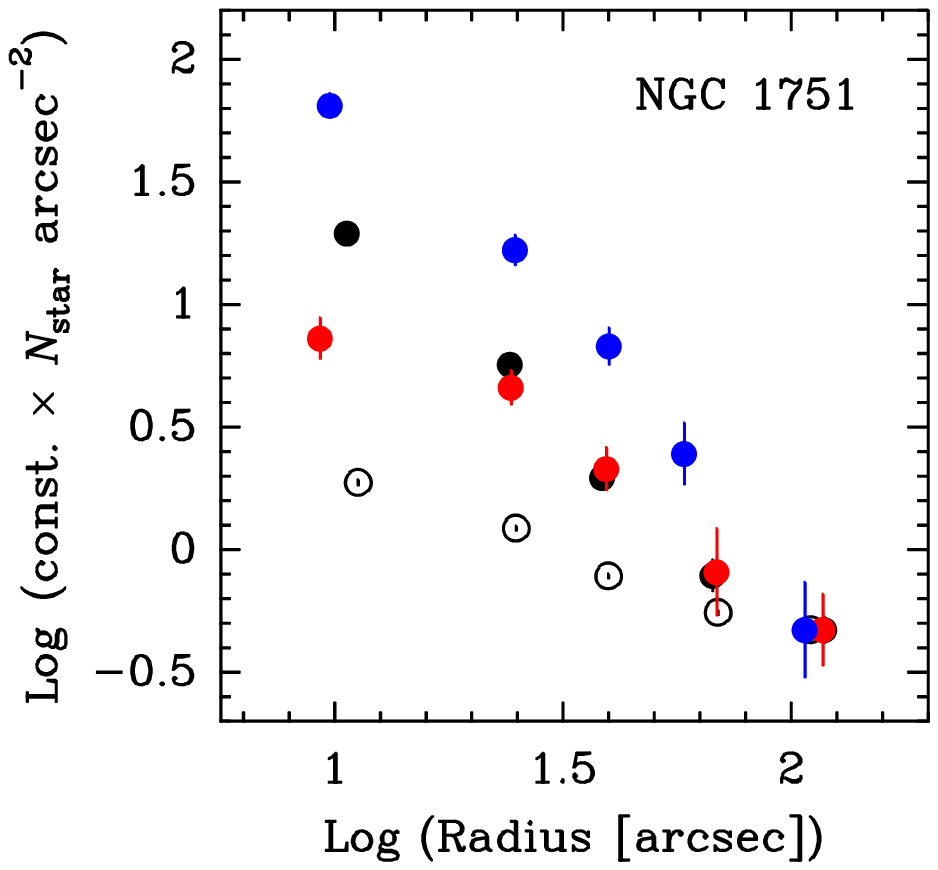,height=5.5cm}
\psfig{figure=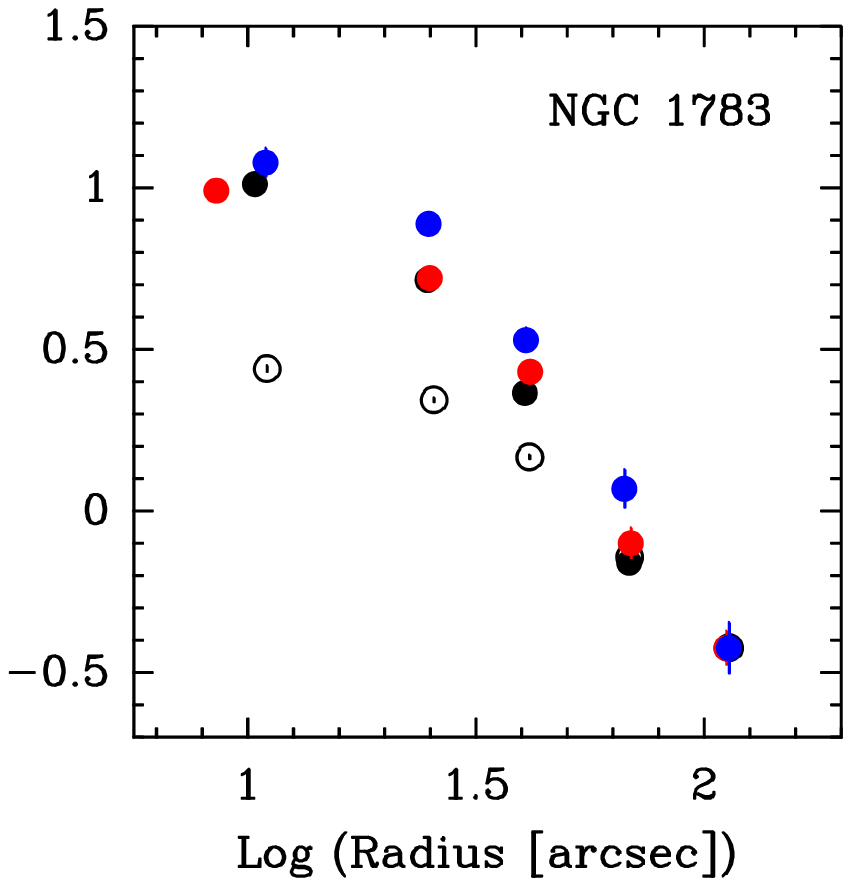,height=5.6cm}
\psfig{figure=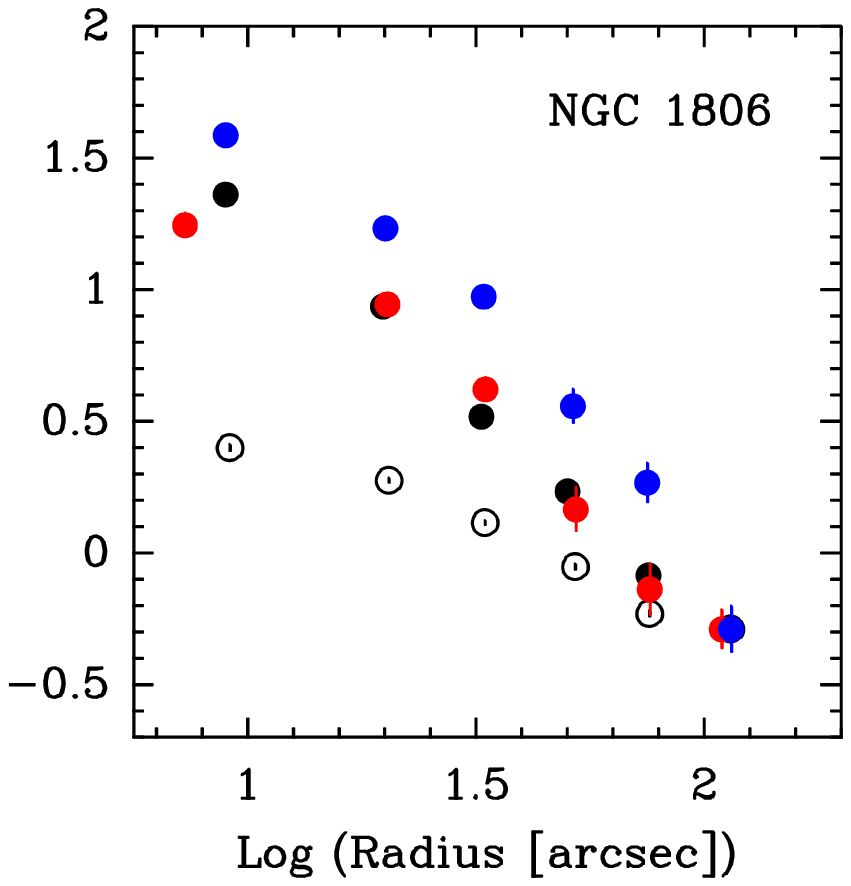,height=5.6cm}
}
\vspace*{1mm}
\centerline{
\psfig{figure=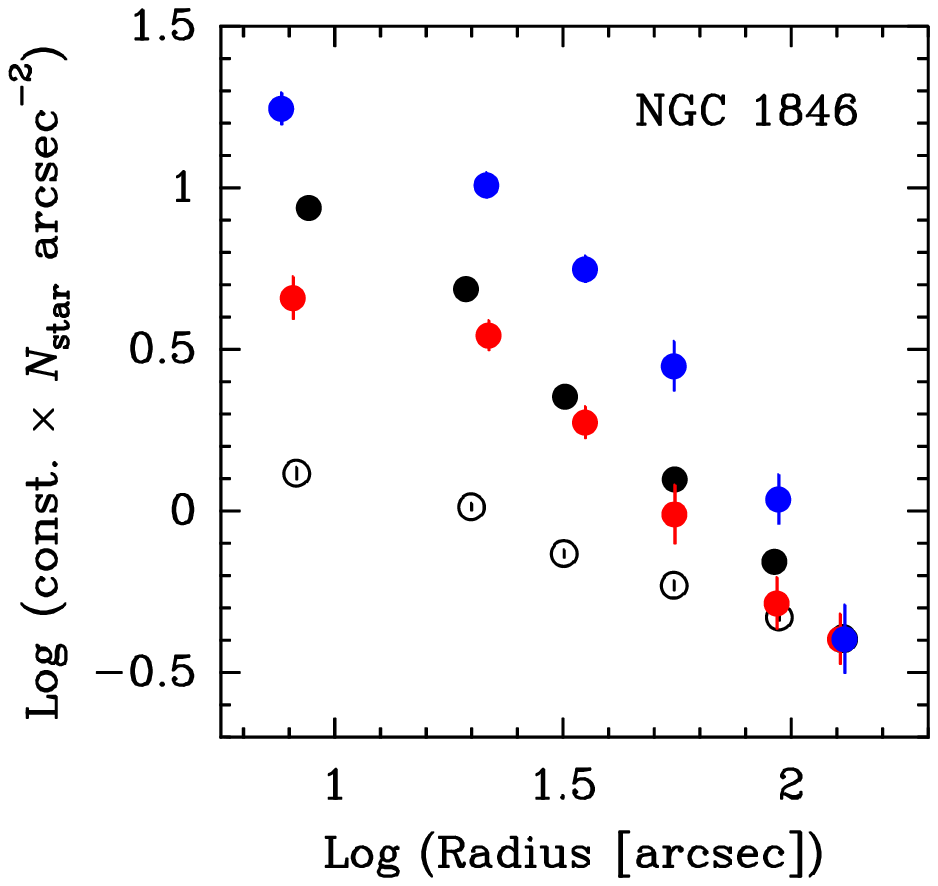,height=5.6cm}
\psfig{figure=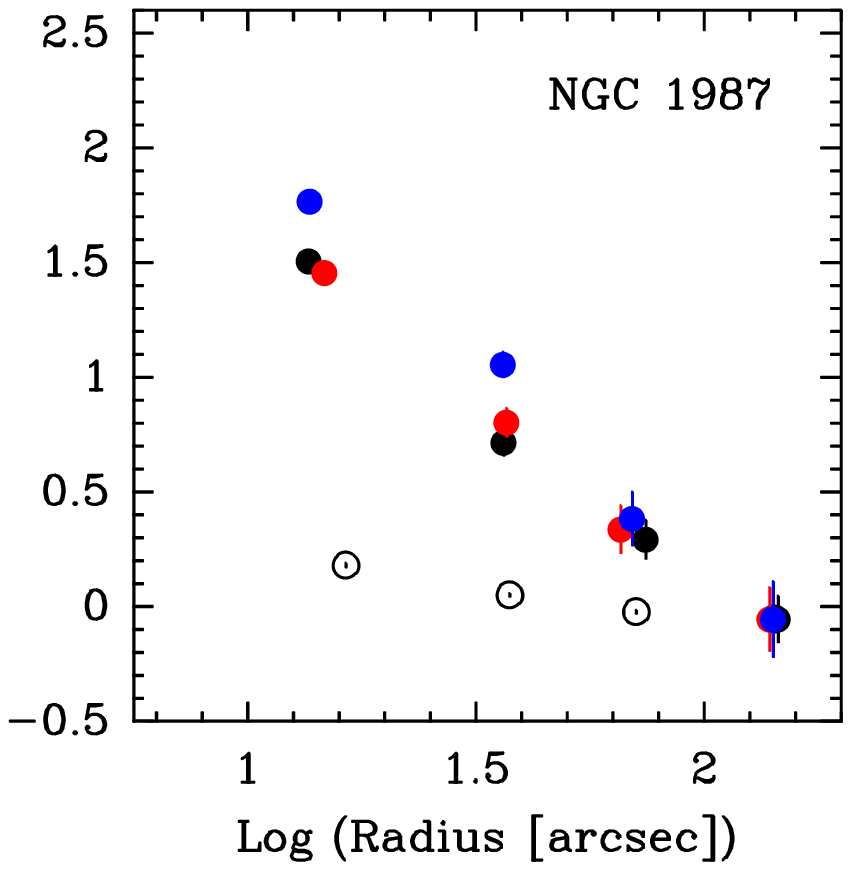,height=5.5cm}
}
\vspace*{1mm}
\centerline{
\psfig{figure=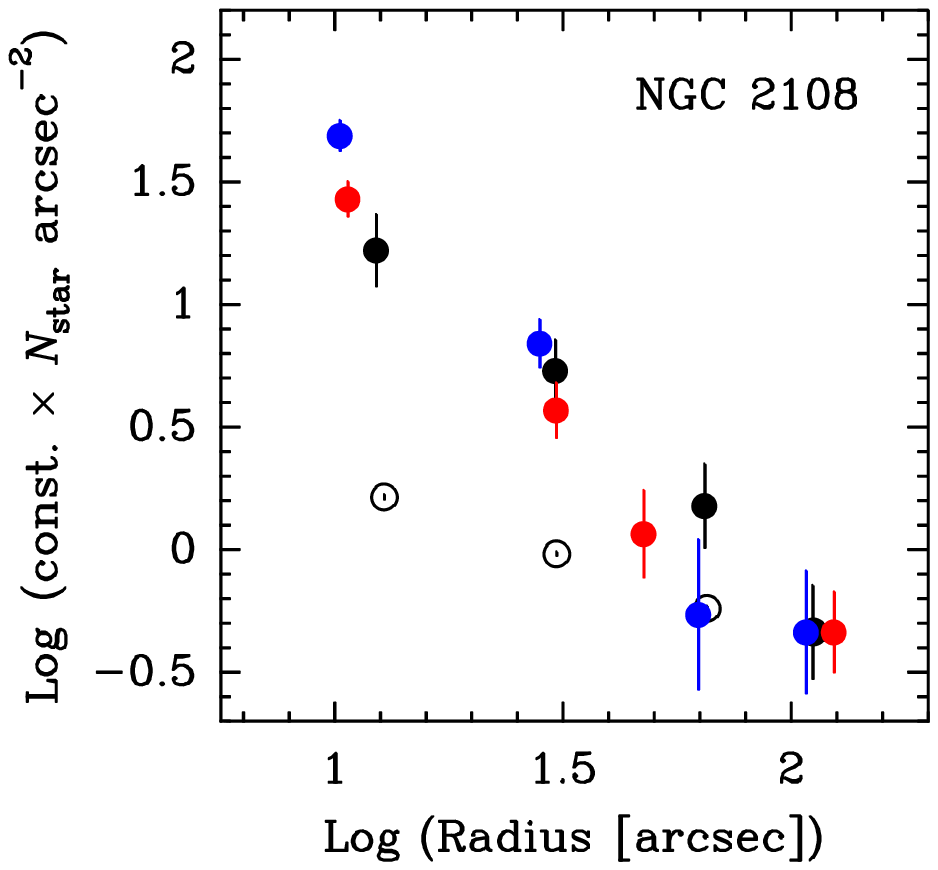,height=5.5cm}
\psfig{figure=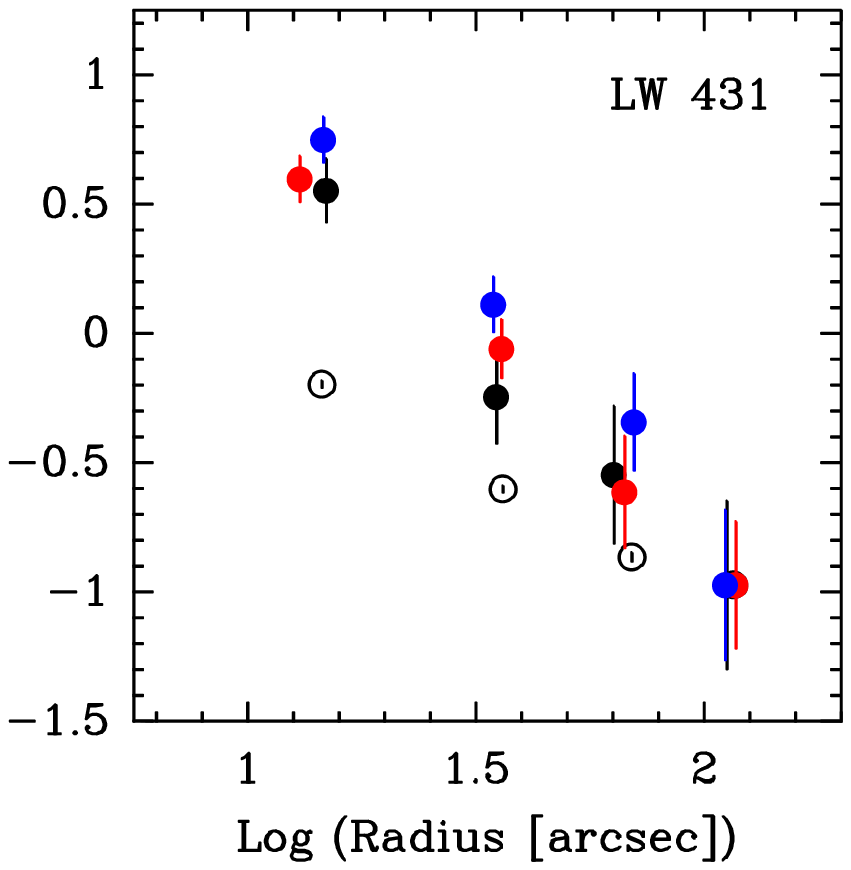,height=5.5cm}
}
\caption{Radial surface number density distributions of stars in different
  evolutionary phases, as illustrated in the CMD in Figure
  \ref{f:regions} with solid lines. Cluster names are given in the
  top right in each panel. Open circles:\ all stars in the CMD. Black  
  circles:\ Stars in the RC, RGB, and AGB. Red circles: Lower MSTO
  region. Blue circles:\ Upper MSTO region. 
  The absolute surface number density values on the Y axis refer to the open 
  circles (``all stars''). The surface number densities of the
  other star types are normalized to that of ``all stars'' at the outermost
  radial bin. See discussion in \S\ \ref{s:rad_dist}.
\label{f:rad_dists}}
\end{figure*}

We derived radial distributions of the following regions on the CMDs
of the seven star clusters in our sample that contain eMSTOs: 
{\it (i)\/} `all' stars in the CMD, {\it (ii)\/} the `upper half' of
the eMSTO region, {\it (iii)\/} the 'lower half' of 
the eMSTO region, and {\it (iv)\/} the Red Clump (RC) stars and the RGB and
AGB stars above the RC. Note that regions {\it (ii) -- (iv)\/} were found
in Paper II to have less than 5\% contamination by field stars for every
cluster. Fig.\ \ref{f:regions} depicts these regions on the CMD of
NGC~1806 with solid lines, while the completeness-corrected radial
distributions of stars in these regions are shown in Fig.\
\ref{f:rad_dists}. The size of radial bins in all panels of Fig.\
\ref{f:rad_dists} was determined by the requirement that there be at
least 30 stars in the outermost radial bin for any of the CMD regions
considered.  

Fig.\ \ref{f:rad_dists} shows several items of interest, 
discussed in turn below. 
\vspace*{-1.8ex}
\begin{description}
\item[(i)] For many of the brighter (more massive) clusters in
  our sample, the stars in the upper (brighter) half of the eMSTO
  region are significantly more centrally concentrated than the stars
  in the lower (fainter) half of the eMSTO. In fact, 
  the stars in
    the upper half of the eMSTO show a central concentration that is
    even stronger than that of (more massive) stars in the upper
    RGB/AGB in the clusters NGC~1751 and NGC~1846 (and, to a somewhat
    lesser extent, NGC 1806). This means that mass segregation cannot
    be the cause of these differences. The radial distribution of the
    RGB/AGB stars in these three massive clusters typically follows
    that of the lower half of the eMSTO except for the innermost few
    radial bins where it is intermediate between the radial
    distributions of the upper and lower halves of the eMSTO region. 
\item[(ii)] Interestingly, this difference in central
  concentration between the upper and lower halves of the eMSTO is not
  quite as strong for NGC~1783, the most luminous 
  cluster in our sample. 
  The case of NGC~1783 is discussed in some detail in \S\,\ref{s:vesc}.4 below. 
\item[(iii)] The difference in central
  concentration between the upper and lower halves of the eMSTO is
  insignificant for the lower-luminosity clusters LW\,431,
  NGC~1987, and NGC~2108.   
\item[(iv)] Finally, the radial gradient of `all' stars in the
  clusters (open circles in Fig.\ \ref{f:rad_dists}) is significantly
  shallower than that of the MSTO and RGB/AGB populations. This mainly
  reflects the relatively strong contamination by LMC field stars in
  the lower MS region of the CMD of these clusters (see Paper II).
\end{description}
We conclude that the upper and lower parts of the eMSTOs in (at least
some of) these star clusters correspond to {\it intrinsically physically
  different populations\/} which seem to have experienced different amounts of
violent relaxation during their collapse and/or different dynamical evolution
effects. 

We believe that this finding is relevant in the context of the two main
interpretations of the eMSTO regions in the recent literature: {\it
  (i)\/} bimodal age distributions or spreads in age
(\citealt{mack+08,milo+09,gira+09}; Paper I; Paper II) and {\it (ii)\/}
spreads in rotation velocity among turnoff stars (\citealt{basdem09}, but see
\citealt{gira+11}). Specifically, if the width of the 
eMSTO in intermediate-age star clusters is mainly caused by one or more secondary
generation(s) of stars having formed within the cluster from material shed in
slow winds of stars of the first generation and retained by the cluster, the  
simulations of \citet{derc+08} show that the 
younger generation (i.e., the {\it upper\/} half of the eMSTO region) would be
more centrally concentrated than the older one if the age of the cluster is
less than or similar to the half-mass relaxation time (which is the
case for the four most massive clusters in our sample, cf.\
Table~\ref{t:dynamics}). Conversely, if the width of the eMSTO is mainly due
to a range in rotation speeds among stars with masses $1.2 \la
{\cal{M}}/M_{\odot} \la 1.7$ \citep{basdem09}, one might expect more rapid
rotators (i.e., the {\it lower\/} half of the eMSTO region according to
\citealt{basdem09}) to be initially more centrally concentrated than
less rapid rotators 
 since observational studies have found young stars in dense
  star clusters to show higher rotation rates than similar stars in
  the field and hence born in presumably less dense stellar aggregates
  \citep[e.g.,][]{kell04,stro+05}.  
While detailed modelling of the dependencies of the radial 
distribution of stars in star clusters on stellar rotation velocity
and cluster age is still lacking, the observation that the upper
(i.e., brighter) half of the eMSTO population is significantly more
centrally concentrated than the lower (i.e., fainter) half in several
star clusters in our sample seems to indicate that age effects {\it
  are\/} responsible for the broadening of the eMSTO (at least in
those clusters).  

\subsection{Trends with Cluster Escape Velocity as a Function of
  Time} \label{s:vesc} 

In a scenario where secondary generations form from
material shed in slow stellar winds of the first generation, expected
wind speeds for intermediate-mass AGB (IM-AGB) stars are about 10\,--\,20 \kms\
\citep[e.g.,][]{vaswoo93,mars+04}. For fast-rotating massive stars
(FRMS), wind speeds range between about ten to a few hundreds of \kms\
\citep{port96,porriv03,wuns+08}. For massive interacting binary stars,
one can expect wind velocities of a few tens of \kms\ as well:
Observations of the well-studied system RY Scuti have 
shown ejection velocities of 30\,--\,70 \kms\ \citep{smit+02,demink+09}. 
If any (or all) of these types of stars provide the material for secondary
generations in star clusters, one would expect the ability of star clusters to
retain this material to scale with their escape velocities at the time such
stars are present in the cluster (i.e., at ages of $\sim$\,5\,--\,30 Myr for the
massive stars and $\sim$\,50\,--\,200 Myr for IM-AGB stars). 
Furthermore, 
 a well-defined relation has been suggested to exist
  between the most massive star in a cluster and the initial 
 cluster cloud mass in that initially more massive clusters host more
 massive stars. The physical effect that causes this relation may be 
 that retention of feedback energy from more massive stars requires
 more massive gas clouds \citep{weikro06}.  
Hence, if massive stars are a significant source of the material used
to form secondary generations in clusters, one would expect the
existence of a relation between the fraction of second-generation
stars and the mass of the cluster.  

With this in mind, we estimate masses and escape velocities of the clusters in
our sample for both the present time and at an age of 10$^7$ yr (i.e., after the
cluster has survived the era of gas expulsion and violent relaxation, see
e.g.\ \citealt{baum+08}) as follows. For comparison with clusters in the same
age range that do {\it not\/} contain eMSTO regions, we also do these
calculations for the four LMC clusters NGC~1644, NGC~1652, NGC~1795, and
IC~2146 from the study of \citet{milo+09}.  

\subsubsection{Present-day Masses and Half-Mass Radii}

Present-day masses and half-mass radii for the clusters in our sample are
adopted from Paper II (see its Tables 1 and 3), using ${\cal{M}}/L_V$ ratios
of the SSP models of \citet{bc03}. For the four clusters found by
\citet{milo+09} not to contain eMSTOs, we use the total $V$-band
magnitudes in \citet{bica+96} along with age, distance, and foreground
reddening values from \citet{milo+09}. Since we cannot find half-light radii
published for those four clusters in the literature, we assume an average
value of clusters in the LMC within the age range 1\,--\,2 Gyr from the
literature (\citealt{macgil03}; Paper II) along with a suitably large
uncertainty, namely $r_{\rm h} = 8 \pm 4$ pc.  

\subsubsection{Early Cluster Mass Loss}

We make a distinction between model clusters with and without initial mass
segregation, the inclusion of which can have a strong impact on cluster
mass loss and dissolution due to the strong expansion of the cluster in
response to rapid mass loss associated with supernova (SN) type II explosions
of first-generation stars \citep[e.g.,][]{chewei90,fukheg95,vesp+09}. To
properly represent the case of clusters with initial mass segregation for
this paper, a selection needs to be made as to the initial properties of the
cluster since the recent literature shows that different initial properties
can lead to a wide range of of cluster dissolution times
\citep{baukro07,baum+08,derc+08,vesp+09}. We consider the evolution of cluster
mass and half-mass radius in the simulation called SG-C30 in
\citeauthor{derc+08} (\citeyear{derc+08}; Vesperini 2011), which involves a
tidally limited model cluster that features a moderate degree of initial mass
segregation. This simulation is selected for the following reasons: 
{\it (i)\/} mass-segregated clusters of this type can survive the rapid 
SN-driven mass loss era \citep{vesp+09}; {\it (ii)\/} it yields a
ratio of first-to-second-generation stars of $\sim$\,1:1 at an age of 2 Gyr
\citep[see Fig.~16 of][]{derc+08}, similar to (though somewhat lower
than) what has been found for the more massive clusters in our sample
(see Paper II; \citealt{milo+09}). Note that simulations of clusters with
initially more strongly mass-segregated stellar distributions can yield
results similar to that of the SG-C30 simulation if the cluster does not fully
fill its Roche lobe \citep{vesp+09}. We also account for the effect of mass
loss due to stellar evolution, most of which takes place in the
first $10^8$ yr. The evolution of this mass loss is derived using the
\citet{bc03} SSP models using the \citet{chab03} IMF\footnote{Use of the
  \citet{salp55} IMF results in initial cluster masses that are
  $\simeq$\,9\% lower than those calculated here.}. The fast stellar
evolution-driven mass loss in the first 10$^8$  yr, 
 in particular that due to feedback and supernovae from massive O and B-type 
  stars, 
also causes an expansion of the cluster. This is accounted
for  by following the prescriptions of \citet{hills80} assuming
adiabatic expansion, since the crossing time of star clusters is much
shorter than the time scale for stellar evolution-driven mass loss
\citep[see also][]{baum+08}.   

\subsubsection{Long-term Cluster Mass Loss}

\begin{figure}[tb]
\centerline{\includegraphics[width=8.3cm]{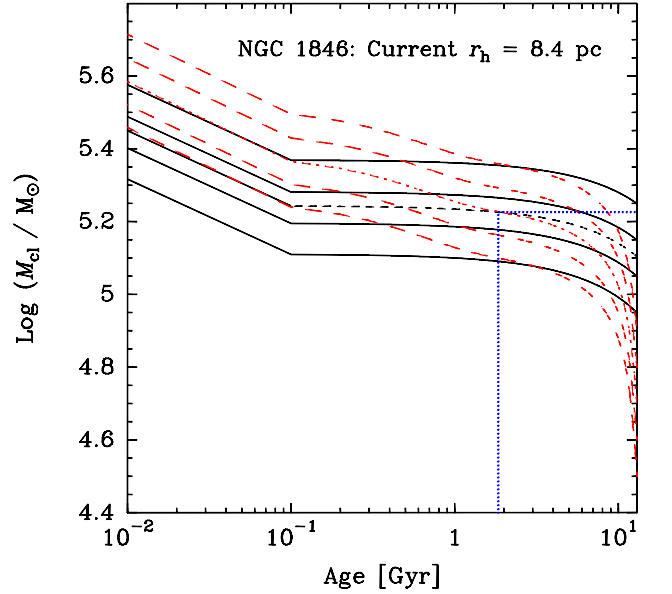}}
\caption{Illustration of the method used to calculate cluster masses as
  function of age. (Black) solid lines and the short-dashed line represent
  calculations for initially non-mass-segregated clusters, whereas
  (red) long-dashed lines and the (red) dashed-dotted line represent
  calculations for an initially mass-segregated cluster model (see discussion
  in \S\ \ref{s:vesc}). Four model calculations are shown for the current
  half-mass radius of NGC~1846 and spaced by log(${\cal{M}}_{\rm
    cl}/M_{\odot}$) = 0.1 at an age of 13 Gyr. The short-dashed and
  dashed-dotted lines depict the adopted mass evolution for NGC~1846. 
  The (blue) dotted lines indicate the current age and mass of NGC~1846. 
\label{f:massevol}}
\end{figure}

At ages $> 100$ Myr we account for long-term dynamical cluster destruction
mechanisms and follow \citet{fz01} and 
\citet{mclfal08} who find that the change in shape of the star cluster mass
function from young ($\la 5 \times 10^8$ yr) to old ($\ga 10^{10}$ yr)
systems is mainly due to mass-density-dependent evaporation (also often
called ``two-body relaxation''). We use the rate of mass loss 
due to evaporation derived by \citet[][their Eq.\ (5)]{mclfal08}: 
\begin{equation}
\overline{\mu_{\rm ev}} \simeq 1100 \left(
  \frac{\rho_{\rm h}}{M_{\odot}\,{\rm pc}^{-3}} \right)^{1/2} M_{\odot} \, 
  {\rm Gyr}^{-1}
\end{equation}
where $\overline{\mu_{\rm ev}}$ is the mass loss rate due to evaporation averaged
over a cluster lifetime and $\rho_{\rm h} \equiv 3\,
{\cal{M}}_{\rm cl}/8 \pi r_{\rm h}^3$ is the half-mass density of the star cluster. 
The evolution of cluster masses is then calculated for an 
appropriately large grid of masses and half-mass radii at an age of 13
Gyr, both with and without initial mass segregation (implemented as described
in the previous paragraph). 
Cluster mass as a function of time for individual star clusters in our sample
are evaluated by means of linear interpolation between grid lines using masses 
at the mean age of the cluster in question. 
The process of calculating cluster masses as a function of time 
is illustrated in Fig.\ \ref{f:massevol} for the case of NGC~1846. For
a typical cluster in our sample, the difference in initial cluster
mass between mass-segregated and non-mass-segregated 
star clusters is of order $\Delta\log {\cal M}_{\rm cl} \approx 0.1-0.15$ dex,
with the initially mass-segregated clusters losing a larger fraction of
their initial mass. 

\subsubsection{Escape Velocities}

Escape velocities required to reach the tidal radius of the cluster are
calculated as follows: 
\begin{equation}
v_{\rm esc} (t) = f_{\rm c}\,\sqrt{\frac{{\cal{M}}_{\rm cl}(t)}{r_{\rm
      h}(t)}} \; \; \mbox{km s}^{-1}
\end{equation}
where ${\cal{M}}_{\rm cl} (t)$ is the cluster mass in \Msun\ at time $t$,
$r_{\rm h, t} (t)$ is the cluster's half-light radius in pc at time $t$, and
$f_{\rm c}$ is a coefficient that takes the dependence of $v_{\rm esc}$
on the concentration index $c$ of \citet{king62} models into
account. Values for  $f_{\rm c}$ are taken from \citet{geor+09}. For
convenience, we define $v_{\rm esc, 7} \equiv v_{\rm esc}
(10^7\;\mbox{yr})$ hereinafter.  

To parametrize the relative central concentrations of the upper and 
lower half of the eMSTO region in the star clusters, we define a ``relative
concentration parameter'' as follows: 
\begin{equation}
C_{\rm rel} \equiv \frac{(n_{\rm core, upp}-n_{\rm bck, upp})}{n_{\rm bck, upp}} 
  \left/ \frac{(n_{\rm core, low}-n_{\rm bck, low})}{n_{\rm bck, low}} \right. 
\label{eq:Crel}
\end{equation}
where $n_{\rm core, upp}$ and $n_{\rm bck, upp}$ are surface number densities
of stars in the upper half of the eMSTO region within the King core radius and
the background region, respectively, and $n_{\rm core, low}$ and 
$n_{\rm bck, low}$ are the same parameters for stars in the lower half of the
eMSTO region. The background region was defined as the outermost radial bin
used for the plots in Figure~\ref{f:rad_dists}. Values for ${\cal{M}}_{\rm
  cl}$, $r_{\rm h}$, $t_{\rm relax}$, $v_{\rm esc}$, and $C_{\rm rel}$ are given in
Table~\ref{t:dynamics}, both for the present time and at an age of 10$^7$
yr. We adopt an uncertainty of 20\% for present-day cluster mass values and
use that in calculating uncertainties of $t_{\rm relax}$ and $v_{\rm esc}$. 

\begin{figure*}[tbh]
\centerline{
\includegraphics[height=8.cm]{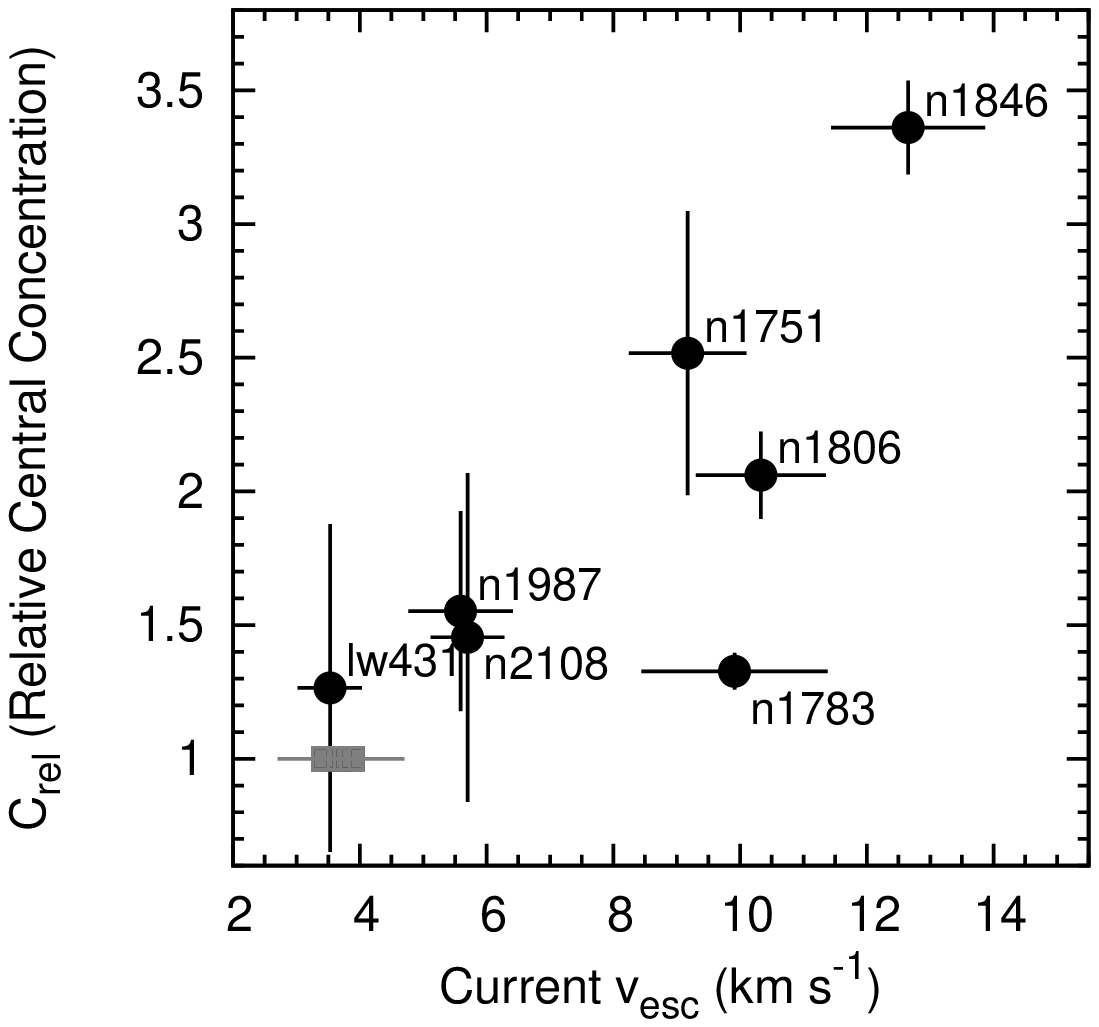}
\includegraphics[height=8.cm]{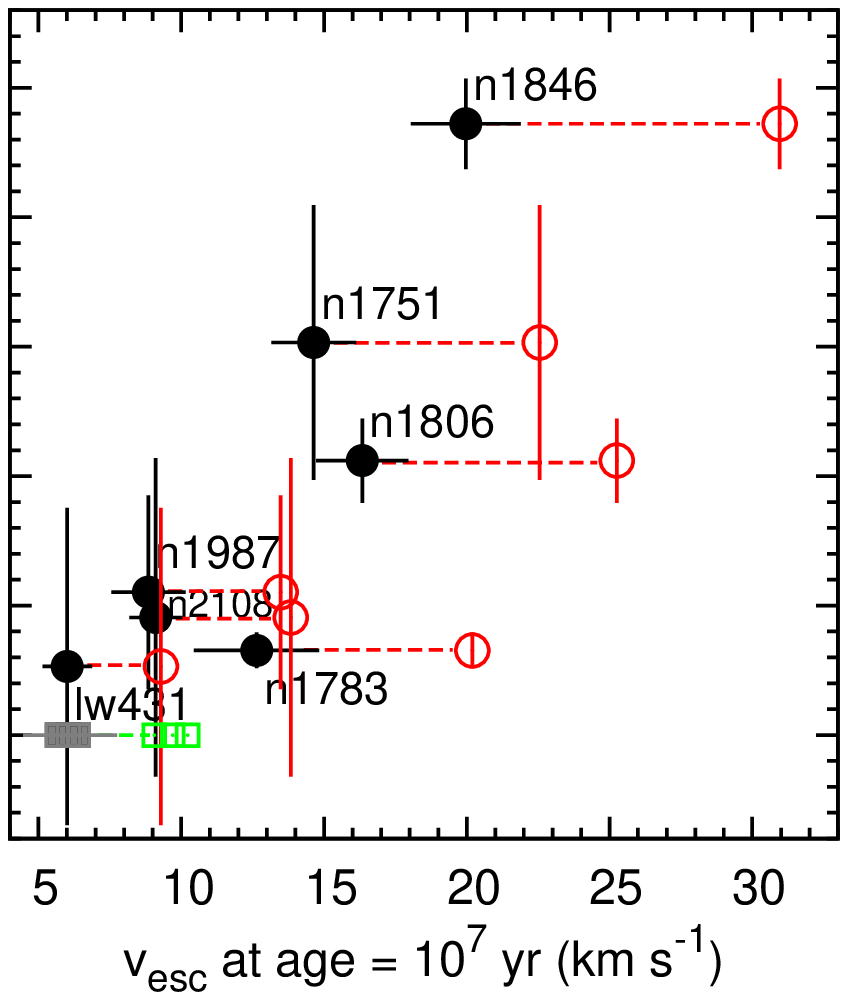}
}
\caption{{\it Left panel}: Parameter $C_{\rm rel}$, a measure of the
  relative central concentration of the upper versus the lower half of the
  MSTO region (cf.\ Eq.\ \ref{eq:Crel}), plotted against {\it current\/} escape
  velocity. {\it Right panel}: $C_{\rm rel}$ against escape velocity at an age
  of 10$^7$ yr. Filled circles represent data for clusters in our
  sample using a model without initial mass segregation while open (red)
  circles do so using a model involving initial mass segregation. 
  Clusters shown by \citet{milo+09} {\it not\/} to exhibit eMSTO regions
  are represented by filled (grey) squares and open (green) squares for 
  models without and with initial mass segregation, respectively. The names of
  the star clusters in our sample are indicated to the upper or lower right of
  their data points. See discussion in \S\ \ref{s:vesc}.  
\label{f:conc_vs_vesc}}
\end{figure*}

$C_{\rm rel}$ is plotted versus cluster $v_{\rm esc}$ at the current age and
at an age of 10$^7$ yr in the two panels of Figure~\ref{f:conc_vs_vesc}. By
definition, we set $C_{\rm rel} \equiv 1$ for the clusters found by
\citet{milo+09} {\it  not\/} to host eMSTOs. 
The left panel shows that the clusters with the highest current escape
velocities tend to have the highest value of $C_{\rm rel}$, although the
current escape velocities are barely high enough to retain material from even
the slowest reported winds of IM-AGB stars (i.e., $\sim$\,10 \kms). However,
the escape velocities were significantly higher at an age of 10$^7$ yr as
illustrated by the right panel of Figure~\ref{f:conc_vs_vesc}. Moreover, the
clusters with $v_{\rm esc, 7} \ga 15$ \kms\ have significantly higher values of
$C_{\rm rel}$ than  those with $v_{\rm esc, 7} \la 10$ \kms, and the
correlation between $C_{\rm rel}$ and $v_{\rm esc}$ becomes 
stronger than in the left panel. These trends are present whether or not one
takes initial mass segregation of stars into account in the calculations of
$v_{\rm esc, 7}$.    Summarizing these results, it appears that intermediate-age
star clusters in the LMC with higher (initial) escape velocities have upper
(``younger'') MSTO stars that are more centrally concentrated
than their counterparts in the lower (``older'') MSTO. This finding is
discussed further in \S\,\ref{s:disc} below. 
  The one cluster that seems to deviate slightly from the trends
  mentioned above and shown in Figure~\ref{f:conc_vs_vesc} is
  NGC~1783. This is the cluster with the 
  the largest radius among our sample. In fact, its radial extent is
  so large that the field of view of the HST/ACS image is too small to
  allow a robust determination of its King concentration parameter
  (see Paper II), which translates into a relatively large uncertainty
  of its half-light radius and hence of its escape velocity. Future
  determinations of its $r_h$ from wider-field imaging should help
  improve the accuracy of its escape velocity and hence its
  position in Figure~\ref{f:conc_vs_vesc}.

\section{Implications Regarding The Nature of Multiple Populations in Star
  Clusters} \label{s:disc} 

The results described above have implications regarding the nature of
multiple stellar generations in star clusters in general, 
including the situation seen in many (ancient) globular clusters in our
Galaxy. To constrain the scope of this broad topic somewhat, we restrict the
following discussion to the case of star clusters that were never massive
enough to retain gas expelled by energetic supernova (SN) explosions
and/or to capture significant numbers of field stars (e.g., from their
host dwarf galaxies). In practice this restriction corresponds roughly to
the exclusion of ancient star clusters with current masses $\ga 2
\times 10^6$ \Msun\ \citep[cf.][]{basgoo06,fell+06} at an age of 13
Gyr. The LMC star clusters studied in this paper are and were indeed
always less massive than that limit.  

Star clusters in our sample with initial escape velocities
$v_{\rm esc, 7} > 15$ \kms\ have $C_{\rm rel}$ values that are
significantly higher than unity and correlate with $v_{\rm esc, 7}$. Hence we
postulate that the extended morphology of their MSTO is {\it mainly caused by a
  range in age}, although we cannot formally rule out that the
widening of the MSTO is partly due to an additional physical process
for those clusters as well.  

As to the star clusters in our sample with initial escape velocities $v_{\rm
  esc, 7} \la 10$ \kms, they have $C_{\rm rel}$ values consistent with
unity. Under the assumption that the material used to form
secondary generations of stars within clusters is mainly produced
internally from stellar winds with $v \ga 10 - 15$ \kms, it seems
plausible that eMSTO regions in such clusters may be due in part to
physical processes unrelated to a range in age, e.g., one or more processes
that are independent of (or only slightly dependent on) the location of stars
within the cluster. This could be a range of stellar rotation velocities as
proposed by \citet{basdem09} and/or other (yet unidentified) physical
processes, possibly related to interactions between close binary stars since
the binary fractions are significant (15\,--\,35\%, cf.\ Paper II) in
the clusters studied here. Note however that the fact that the
clusters found by \citet{milo+09} {\it not\/} to harbor eMSTOs are all
at the low end of the mass (and $v_{\rm esc, 7}$) range of the
clusters considered here seems inconsistent with the stellar rotation
scenario in that one would not a priori expect the distribution of
stellar rotation rates to depend on cluster mass or escape velocity. 
In this context we note that an age range would still be able to
account for a widening of the MSTO in low-mass clusters in our
sample in case their escape velocities at an age of 10$^7$ yr turn out
to be higher than our estimates. This would, for example, be the case
if the initial mass segregation of such clusters was stronger than in
the model we used in \S\,\ref{s:vesc}.2. In that case, the low values 
of $C_{\rm rel}$ in these low-mass clusters can still be explained by their
short half-mass relaxation times which are significantly shorter than their
age (cf.\ Table~\ref{t:dynamics}) so that initial differences in radial
distribution between generations of stars would now have been eliminated. 

These results impact our understanding of
multiple populations in ancient globular clusters. For
example, the finding of multiple populations and a Na-O
anticorrelation in M\,4 \citep[][and references
therein]{marino+08} was described as surprising given its low mass of $\sim 6
\times 10^4$ \Msun. However, using its current values for half-light radius
$r_{\rm h} = 2.23$ pc and King concentration parameter $c = 39$
\citep{harris96}, an age of 13 Gyr, and the methodology outlined in
\S\,\ref{s:vesc}, we calculate a present-day escape velocity
$v_{\rm esc} = 16$ \kms\ and a $v_{\rm esc, 7} = 52$ \kms\ for M\,4
(assuming no initial mass segregation). This is more than sufficient
to have retained chemically enriched material from slow stellar winds
of the first stellar generation. In fact, for typical half-mass
cluster radii $r_{\rm h} \la 3$ pc, we would expect ancient star
clusters with current masses as low as $\sim 5 \times 10^3$ \Msun\ to
have been able to retain such material (see also \citealt{conr11} for
a scenario based on evolution of cluster masses). 

\begin{figure}[tpb]
\centerline{
\includegraphics[width=8.3cm]{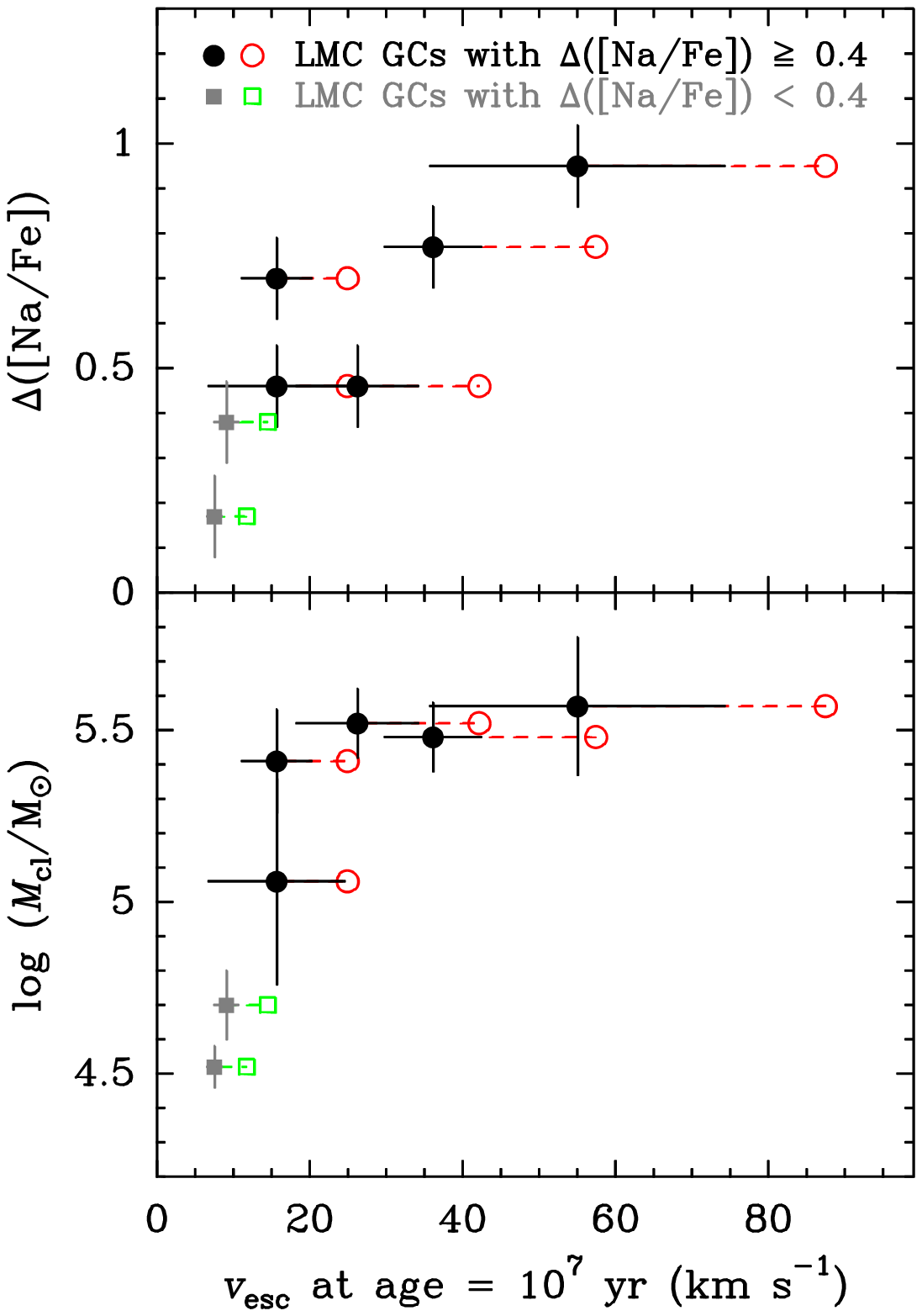}
}
\caption{{\it Top panel}: $\Delta$\,([Na/Fe]), defined as the measured spread
  in [Na/Fe] among stars in LMC clusters studied by
  \citet{mucc+08,mucc+09,mucc+11}, versus cluster escape velocity at an
  age of 10 Myr. Error bars for $\Delta$\,([Na/Fe]) indicate the
  typical uncertainty of individual [Na/Fe] measurements. {\it Bottom
    panel}: Present-day cluster mass against escape velocity at an age 
  of 10$^7$ yr for the same clusters. Filled (black) circles represent
  data for clusters with $\Delta (\mbox{[Na/Fe]}) \ge 0.4$ (a
  4$\sigma$ effect) using a model without initial mass segregation while open (red) 
  circles do so using a model involving initial mass segregation. 
  Clusters with $\Delta (\mbox{[Na/Fe]}) < 0.4$ 
  are represented by filled (grey) squares and open (green) squares for 
  models without and with initial mass segregation, respectively. 
  See discussion in \S\ \ref{s:disc}.  
\label{f:mucc_fig}}
\end{figure}

Similarly, recent results on the presence or absence of significant
light-element abundance variations within LMC clusters can be
understood by considering the clusters' initial escape
velocities. \citet{mucc+09} measured Na and O abundances for RGB stars in
three old metal-poor GCs in the LMC which have current masses in the range
$2-4 \times 10^5$ \Msun\ \citep[see][]{macgil03}, and found clear evidence
for Na-O anticorrelations similar to 
those observed in Galactic GCs. On the other hand, \citet{mucc+08} measured
element abundances in four relatively massive intermediate-age (1\,--\,2 Gyr)
clusters in the LMC and found only small variations of element abundance
ratios within each cluster. However, focusing on their [Na/Fe] measurements,
we notice that two clusters in their sample (NGC 1978 and NGC 2173) exhibit a
range of [Na/Fe] values $\Delta \mbox{[Na/Fe]} \geq 0.4$ dex which is a $\geq
4\sigma$ effect where $\sigma$ is the typical measurement error of
[Na/Fe]\footnote{The negligible variation of [O/Fe] in these clusters found by
  \citet{mucc+08} can be explained by their moderately high metallicities for
  which depletion of O in IM-AGB stars is much smaller than for the much more
  metal-poor ancient GCs in the LMC \citep{vendan09,conr11}.}. 
More recently, \citet{mucc+11} measured element abundances of 14 stars in the
young globular cluster NGC~1866 in the LMC and found no sign of a Na-O
anticorrelation (nor a significant spread in [Na/Fe]). Since they also did not
detect Na-O anticorrelations in the massive intermediate-age clusters in their
2008 paper, they argued for ``a different formation/evolution scenario for the
LMC massive clusters younger than $\sim$\,3 Gyr with respect to the old
ones''. To evaluate whether the presence or absence of significant spreads in
[Na/Fe] in the LMC clusters studied by Mucciarelli et al.\
(\citeyear{mucc+08,mucc+09,mucc+11}) could instead ``simply'' be due
to the clusters' masses and escape velocities at an age of 10 Myr, we
first adopt these clusters' current masses, radii, and ages from the
compilation of \citet{macgil03}. Since NGC~1978 was not included in
that compilation, we adopt its age, [Fe/H] and  
$E(B\!-\!V)$ from \citet{milo+09}, we assume a radius of $r_{\rm h} = 8 \pm 4$
pc (cf.\ \S\,\ref{s:vesc}), and we estimate its current mass from its
integrated $V$ magnitude in \citet{goud+06} and the ${\cal{M}}/L_V$ ratios of
the \citet{bc03} SSP models using the Chabrier IMF. We then estimate
these clusters' masses and escape velocities at an age of 10 Myr using
the methodology described in \S\,\ref{s:vesc}. We plot the observed
spreads in [Na/Fe] (defined here as $\Delta (\mbox{[Na/Fe]})$) from
Mucciarelli et al.\ (\citeyear{mucc+08,mucc+09,mucc+11}) as well as
the current cluster masses versus $v_{\rm esc, 7}$ in
Figure~\ref{f:mucc_fig}. Note that {\it (i)\/} the distinction between
clusters with and without significant ($\sim$\,4$\sigma$) spreads in
[Na/Fe] is found at $v_{\rm esc, 7} \approx 15$ \kms\ at an age of 10
Myr, and {\it (ii)\/} $\Delta (\mbox{[Na/Fe]})$ correlates with
$v_{\rm esc, 7}$, indicating that clusters with higher initial escape
velocities are able to retain material with higher variance of
light-element abundances. This is consistent with
the results shown in  Figure~\ref{f:conc_vs_vesc} for the $C_{\rm rel}$
parameter of the clusters in our sample, and underlines the importance of the
clusters' initial escape velocities in their ability to retain slow wind
material from stars of the first generation. 

\begin{figure*}[tbp]
\centerline{
\includegraphics[width=0.8\textwidth]{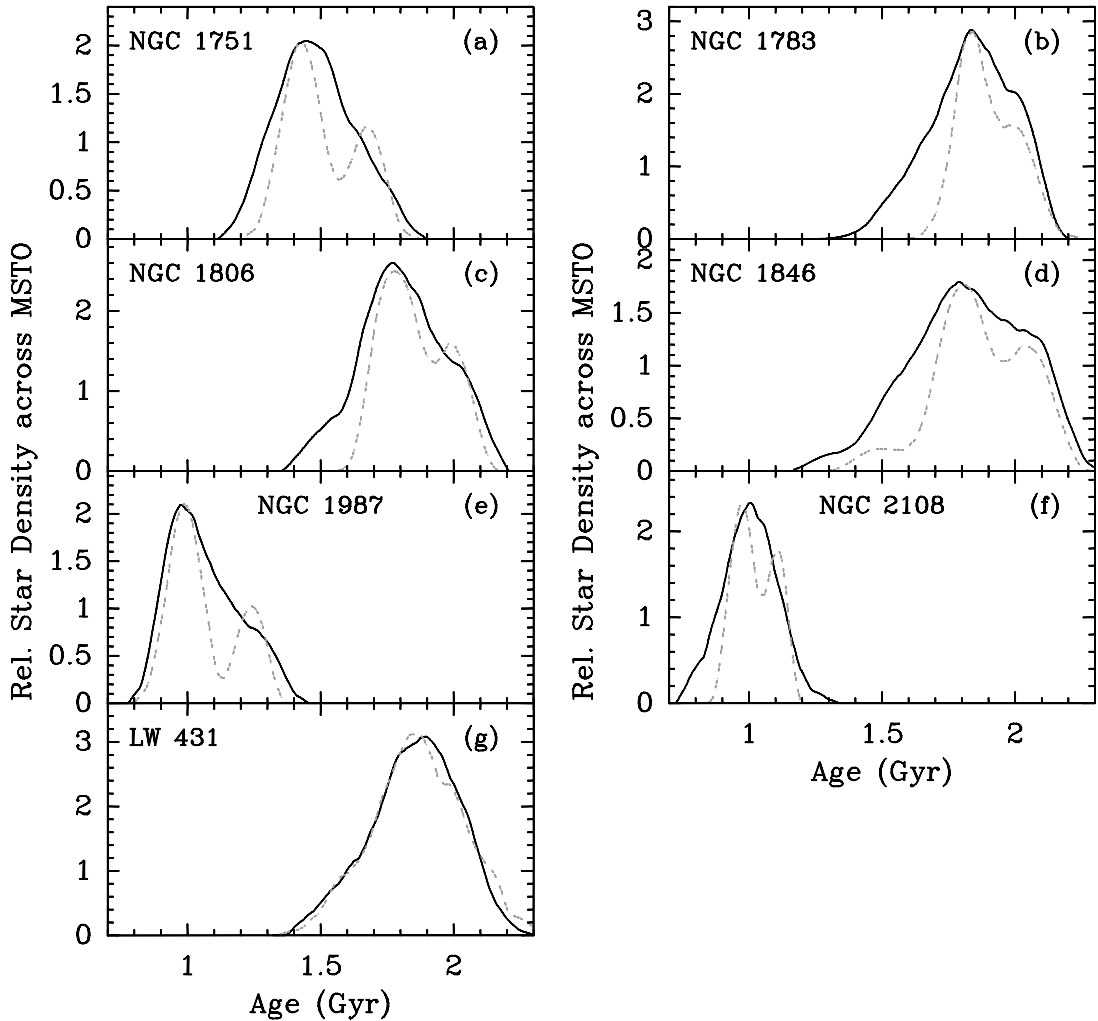}
}
\caption{Age distributions of the clusters in our sample, reproduced from
  Papers I and II, derived from probability density profiles of the number
  density of stars across the eMSTO regions. The solid lines represent cluster
  stars. For comparison, the grey dashed lines represent the best-fit two-SSP
  simulations of the cluster's age distributions. See Paper II for details. 
  The names of the star clusters are indicated in the legend of each panel. See
  discussion in \S\ \ref{s:disc}.  
\label{f:agedists}}
\end{figure*}

Our results for the star clusters
with $v_{\rm esc, 7} \ga 15$ \kms\ are generally
consistent with the conclusions drawn for NGC~1846 in Paper~I
\citep[see also][]{renz08,conspe11}. 
In particular, we believe our results are most consistent with the ``in situ''
scenario \citep[e.g.,][]{derc+08,renz08} in which star clusters with masses
high enough  to retain ejecta in slow winds of stars of the first generation
gather this material in  their central regions where second-generation
star formation can occur. The observed relation between the $C_{\rm
  rel}$ parameter and the initial escape velocity of the cluster
strongly suggests that this is taking place. 
In the context of this scenario, we recall that the hitherto suggested
source(s) of the ejecta are FRMS \citep{decr+07}, massive binary stars
\citep{demink+09}, and IM-AGB stars \citep[e.g.,][]{danven07}. Note that the
ejecta from FRMS and massive binary stars are produced on time scales that are
significantly shorter than those from IM-AGB stars ($\sim 10 - 30$~Myr
versus $\sim 50 - 300$~Myr, respectively; see e.g.\
\citealt{decr+07,vendan08}). 
  I.e., all material created by massive stars that could be used to form second
  generation stars would be available in the cluster by the time IM-AGB stars
  would {\it start\/} their slow mass loss.  
To compare these time scales to the age distributions of the eMSTO
clusters in our sample, we show the latter in Fig.~\ref{f:agedists},
reproduced from Papers I and II. Note that Monte Carlo simulations
described in Paper II show that two  populations with ages separated
by 100\,--\,150 Myr or more would result in observable bimodality in
our age distributions derived from the MSTO photometry. Hence, the
combination of the observed age ranges of 200\,--\,500 Myr, 
  the fact that the material lost from first-generation stars throughout the
  cluster needs some time to accumulate in the central regions to allow
  second-generation star formation to occur, and the absence of clear
bimodality in the age distributions seems to suggest at face value
that FRMS and/or massive binary stars could well be significant
contributors to the enriched material used to produce the second
stellar generation. However, IM-AGB stars also seem likely significant
contributors, since the age distributions of the star clusters with
the highest escape velocities in our sample do typically peak near the
younger end of the observed age range (especially for the more massive
clusters in our sample).  

Our results also have an impact on an argument made against the ``in
situ'' scenario that if indeed the eMSTO phenomenon reflects an
range of ages and the second generation of stars is formed from material
produced by first-generation stars, why are no eMSTOs seen in clusters
with ages similar to the typical spread in ages within eMSTO clusters
(i.e., 100\,--\,500 Myr)? The answer seems to be, at least in part, 
that the escape velocities of clusters in this age range that are
nearby enough to detect the presence of an eMSTO are not high enough to
retain the ejecta of the first-generation stars. Using the
masses and radii derived by the study of \citet{macgil03} for
Magellanic Cloud clusters in that age range and applying our
methodology described in \S\,\ref{s:vesc}, we find that there is only
one such cluster (NGC~1856, log ${\cal{M}}_{\rm cl}/M_{\odot} \sim
4.9$, $r_h \sim 3.2$ pc, log age = 8.12) 
that has a $v_{\rm esc, 7}$ value above 15 \kms, i.e., sufficient to 
retain gas lost by slow stellar winds of the first
generation. Unfortunately, the currently available photometric data
for this cluster are inadequate to properly assess the presence of a
range of ages. \citet{kell+11} also recently made an additional,
statistically based argument suggesting that the absence of eMSTO 
clusters in the age range of 100\,--\,300 Myr is not problematic to 
the ``in situ'' scenario. 

In addition to the ``in situ'' scenario, part of the gas used to form 
second-generation stars may have been accreted from the ambient ISM 
$\ga 10^8$ yr after the first generation was formed
\citep[see][]{conspe11,pflkro09}. This scenario is attractive in that {\it (i)\/} it
addresses the concern that the high fraction of second-generation
stars in GCs seem to require more gas than any suggested type of donor stars
can provide, and {\it (ii)\/} gas from the ambient ISM readily provides the
``dilution'' that is likely needed to create the smooth Na-O anticorrelations
seen in ancient Galactic GCs \citep[e.g.,][]{vendan08,vendan09}. In the
absence of such dilution, one would expect a more discrete distribution of cluster
stars in the [Na/Fe] vs.\ [O/Fe] diagram than that observed
\citep[see, e.g.,][]{carr+09}. However, the
uniformity of massive element abundances (e.g., [Fe/H] and [Ca/H]) among
stars in Galactic GCs (except for the most massive ones like $\omega$\,Cen and
M~54) and the razor-sharp RGB sequences seen in the CMDs of the more massive
LMC star clusters in our sample (cf.\ Paper II) do require that
the spread in [Fe/H] and [Ca/H] in the ambient ISM was very small when the
second-generation stars were formed. Future measurements of [Fe/H] and [Ca/H]
variations among RGB stars in the LMC field population may be
able to address this concern. 
A next generation of theoretical simulations similar to those used by
\citet{derc+08,derc+10} may well be able to shed more light on the expected
relative contributions of the possible donor  
star types after taking into account their respective numbers of stars from
the IMF (see, e.g., \citealt{demink+09}), their typical evolutionary time
scales, the abundance patterns of their ejecta, and the mass evolution of star
clusters containing multiple generations of stars.

\section{Summary and Conclusions} \label{s:conc}

We analyzed dynamical properties of 11 intermediate-age star clusters in the
LMC that have high-precision ages and metallicities measured from recent
HST/ACS photometry. 7 of these 11 clusters contain MSTO regions that are
significantly more extended than expected from measurement errors. 

For the three massive clusters NGC\,1751, NGC\,1806, and NGC\,1846, we find
that radial distributions of stars in the upper (i.e., brighter) half of the
eMSTO region are significantly more centrally concentrated than stars in the
lower (i.e., fainter) half of the eMSTO region as well as the (more 
massive) RGB and AGB. Since this cannot be due to dynamical evolution
of a SSP, we conclude that the upper and lower
MSTO regions in those clusters correspond to intrinsically different
populations which have undergone different amounts of violent
relaxation during their collapse.  
To look for an explanation for this finding, we calculate the evolution of
mass and escape velocity for the clusters in our sample from an age of 
10$^7$ yr to the present time, using a combination of stellar evolution models
and dynamical cluster disruption models. The three star
clusters mentioned above turn out to be the only three clusters in our sample
that have the following two properties: {\it (i)\/} escape velocities $v_{\rm
  esc} \ga 15$ \kms\ at an age of 10$^7$  yr, i.e., large enough to retain
material shed by slow stellar winds of the first generation, and {\it (ii)\/}
half-mass relaxation times that are larger than or equal to their age. 
At least for these clusters, we suggest that the main cause of the presence of
the eMSTO region is a range of stellar ages within the cluster. The
data seem most consistent with the ``in situ'' scenario in which 
secondary generations of stars form within the cluster itself
out of gas ejected by stars of the first stellar generation that feature slow
stellar winds. Viable sources of the 
enriched material are thought to include fast-rotating massive stars, massive
binary stars, and intermediate-mass AGB stars. Element abundance ratios from
high-resolution spectroscopy of individual cluster stars should be very useful
in further constraining the nature of the eMSTO regions in massive
intermediate-age star clusters. In particular, one would expect to see
correlated variations in light element abundances (e.g., N, O, Na) among the
stars in the star clusters with relatively high escape velocities in our
sample, likely in a way similar to the Na-O anticorrelation found in ancient
GCs.

\paragraph*{Acknowledgments.}~We acknowledge stimulating discussions with
Aaron Dotter, Iskren Georgiev, Leo Girardi, Selma de Mink, and Enrico
Vesperini.   
 We gratefully acknowledge the thoughtful comments and suggestions
 of the anonymous referee. 
T. H. P.\ acknowledges support by the FONDAP Center for Astrophysics 15010003 and
BASAL Center for Astrophysics and Associated Technologies PFB-06,
Conicyt, Chile. 
R. C.\ acknowledges support from the National Science Foundation through
CAREER award 0847467. This research was supported in part by the National
Science Foundation under Grant No.\ PHY05-51164. 
Support for {\it HST\/} Program GO-10595 was provided by
NASA through a grant from the Space Telescope Science Institute, 
which is operated by the Association of Universities for Research in
Astronomy, Inc., under NASA contract NAS5--26555.

\clearpage

\begin{turnpage}
\begin{deluxetable*}{@{}lccccccccc@{}}
\tablewidth{0pt}
\tablecolumns{10}
\tablecaption{Adopted dynamical parameters of the eMSTO star clusters studied in this
  paper. 
\label{t:dynamics}}
\tablehead{
 \colhead{} & \multicolumn{2}{|c}{log (${\cal{M}}_{\rm cl}/M_{\odot}$)} &
 \multicolumn{2}{|c}{$r_{\rm h}$} & \multicolumn{2}{|c}{$t_{\rm relax}$} &
 \multicolumn{2}{|c|}{$v_{\rm esc}$} & \colhead{} \\
 \colhead{Cluster} & \colhead{Current} & \colhead{10$^7$ yr} &
 \colhead{Current} & \colhead{10$^7$ yr} & \colhead{Current} & \colhead{10$^7$
   yr} & \colhead{Current} & \colhead{10$^7$ yr} & \colhead{$C_{\rm rel}$} \\
 \colhead{(1)} & \colhead{(2)} & \colhead{(3)} & \colhead{(4)} &
 \colhead{(5)} &  \colhead{(6)} & \colhead{(7)} & \colhead{(8)} &
 \colhead{(9)} & \colhead{(10)} \\ 
}
\startdata                                  
\cutinhead{Case I: Model Without Initial Mass Segregation.}
NGC 1751 & 4.82 $\pm$ 0.09 & 5.07 $\pm$ 0.09 & 
 7.1 $\pm$ 0.6 & 5.0 $\pm$ 0.4 & 1.2 $\pm$ 0.2 & 0.9 $\pm$ 0.1 & 
 \phm{1}9.2 $\pm$ 0.9 & 14.6 $\pm$ 1.5 & 2.52 $\pm$ 0.53 \\
NGC 1783 & 5.25 $\pm$ 0.09 & 5.46 $\pm$ 0.09 & 
 \llap{1}4.7 $\pm$ 3.2 & 9.9 $\pm$ 2.2 & 5.2 $\pm$ 1.8 & 3.8 $\pm$ 1.3 &
 \phm{1}9.9 $\pm$ 1.5 & 12.6 $\pm$ 2.3 & 1.33 $\pm$ 0.07 \\
NGC 1806 & 5.03 $\pm$ 0.09 & 5.27 $\pm$ 0.09 & 
 8.9 $\pm$ 0.6 & 6.3 $\pm$ 0.4 & 2.0 $\pm$ 0.2 & 1.5 $\pm$ 0.2 &
 10.3 $\pm$ 1.2 & 16.3 $\pm$ 1.6 & 2.06 $\pm$ 0.16 \\
NGC 1846 & 5.17 $\pm$ 0.09 & 5.41 $\pm$ 0.09 & 
 8.4 $\pm$ 0.4 & 5.9 $\pm$ 0.3 & 2.1 $\pm$ 0.2 & 1.5 $\pm$ 0.1 & 
 12.7 $\pm$ 1.2 & 20.0 $\pm$ 1.9 & 3.36 $\pm$ 0.18 \\
NGC 1987 & 4.49 $\pm$ 0.09 & 4.73 $\pm$ 0.09 & 
 8.9 $\pm$ 2.2 & 6.2 $\pm$ 1.5 & 1.2 $\pm$ 0.4 & 0.9 $\pm$ 0.3 & 
 \phm{1}5.6 $\pm$ 0.8 & \phm{1}8.8 $\pm$ 1.3 & 1.55 $\pm$ 0.37 \\
NGC 2108 & 4.41 $\pm$ 0.09 & 4.66 $\pm$ 0.09 & 
 7.1 $\pm$ 0.6 & 5.0 $\pm$ 0.4 & 0.8 $\pm$ 0.1 & 0.6 $\pm$ 0.1 &
 \phm{1}5.7 $\pm$ 0.6 & \phm{1}9.1 $\pm$ 0.9 & 1.45 $\pm$ 0.62 \\
\phm{N}LW 431 & 4.00 $\pm$ 0.09 & 4.31 $\pm$ 0.09 & 
 7.2 $\pm$ 1.7 & 5.1 $\pm$ 1.2 & 0.6 $\pm$ 0.2 & 0.4 $\pm$ 0.2 &
 \phm{1}3.5 $\pm$ 0.5 & \phm{1}6.0 $\pm$ 0.9 & 1.27 $\pm$ 0.61 \\
\cutinhead{Case II: Model Involving Initial Mass Segregation.}
NGC 1751 & 4.82 $\pm$ 0.09 & 5.17 $\pm$ 0.09 & 
 7.1 $\pm$ 0.6 & 2.6 $\pm$ 0.2 & 1.2 $\pm$ 0.2 & 0.4 $\pm$ 0.1 & 
 \phm{1}9.2 $\pm$ 0.9 & 22.6 $\pm$ 2.3 & 2.52 $\pm$ 0.53 \\
NGC 1783 & 5.25 $\pm$ 0.09 & 5.58 $\pm$ 0.09 & 
 \llap{1}4.7 $\pm$ 3.2 & 5.5 $\pm$ 1.1 & 5.2 $\pm$ 1.8 & 1.6 $\pm$ 0.6 &
 9.9 $\pm$ 1.5 & 20.2 $\pm$ 3.0 & 1.33 $\pm$ 0.07 \\
NGC 1806 & 5.03 $\pm$ 0.09 & 5.37 $\pm$ 0.09 & 
 8.9 $\pm$ 0.6 & 3.3 $\pm$ 0.2 & 2.0 $\pm$ 0.3 & 0.6 $\pm$ 0.1 &
 10.3 $\pm$ 1.2 & 25.3 $\pm$ 2.5 & 2.06 $\pm$ 0.16 \\
NGC 1846 & 5.17 $\pm$ 0.09 & 5.52 $\pm$ 0.09 & 
 8.4 $\pm$ 0.4 & 3.1 $\pm$ 0.1 & 2.1 $\pm$ 0.3 & 0.7 $\pm$ 0.1 & 
 12.7 $\pm$ 1.2 & 31.0 $\pm$ 3.0 & 3.36 $\pm$ 0.18 \\
NGC 1987 & 4.49 $\pm$ 0.09 & 4.82 $\pm$ 0.09 & 
 8.9 $\pm$ 2.2 & 3.3 $\pm$ 0.8 & 1.2 $\pm$ 0.5 & 0.4 $\pm$ 0.1 & 
 \phm{1}5.6 $\pm$ 0.8 & 13.5 $\pm$ 2.0 & 1.55 $\pm$ 0.37 \\
NGC 2108 & 4.41 $\pm$ 0.09 & 4.75 $\pm$ 0.09 & 
 7.1 $\pm$ 0.6 & 2.7 $\pm$ 0.2 & 0.8 $\pm$ 0.1 & 0.3 $\pm$ 0.1 &
 \phm{1}5.7 $\pm$ 0.6 & 13.5 $\pm$ 2.0 & 1.45 $\pm$ 0.62 \\
\phm{N}LW 431 & 4.00 $\pm$ 0.09 & 4.41 $\pm$ 0.09 & 
 7.2 $\pm$ 1.7 & 2.7 $\pm$ 0.6 & 0.6 $\pm$ 0.2 & 0.2 $\pm$ 0.1 &
 \phm{1}3.5 $\pm$ 0.5 & \phm{1}9.3 $\pm$ 1.3 & 1.27 $\pm$ 0.61 \\
\enddata
\tablecomments{Column (1): Name of star cluster. (2): Logarithm of adopted current
  cluster mass (in solar masses). (3): Logarithm of adopted cluster mass at
  an age of 10 Myr. (4): Current cluster half-mass radius in pc. (5): Adopted
  cluster half-mass radius at an age of 10 Myr. (6): Current cluster half-mass
  relaxation time in Gyr. (7): Cluster half-mass relaxation time at an age of
  10 Myr. (8): Current cluster escape velocity at tidal radius in
  \kms. (9): Cluster escape velocity at tidal radius at an age
  of 10 Myr.  (10): Value of $C_{\rm rel}$, defined in \S\,\ref{s:vesc}.4.
  }
\end{deluxetable*}
\end{turnpage}

\end{document}